\documentclass[11pt]{article}
\pdfoutput=1

\usepackage{amsmath,amsthm}
\usepackage{preprint}

\usepackage[utf8]{inputenc}
\usepackage[T1]{fontenc}
\usepackage{graphicx}
\usepackage{float}
\usepackage{booktabs}
\usepackage{array}
\usepackage{hyperref}
\usepackage{cleveref}
\usepackage{xcolor}
\usepackage{tikz}
\usetikzlibrary{arrows.meta,positioning,shapes.geometric,fit,calc,backgrounds}
\usepackage{listings}
\usepackage{enumitem}
\usepackage{caption}
\usepackage{longtable}

\newcolumntype{P}[1]{>{\raggedright\arraybackslash}p{#1}}

\lstdefinelanguage{Isabelle}{
  morekeywords={locale,fixes,assumes,shows,lemma,theorem,definition,
    interpretation,datatype,fun,where,record,type_synonym,by,auto,
    cases,metis,unfold_locales,rule,proof,qed,imports,theory,begin,end},
  sensitive=true,
  morecomment=[s]{(*}{*)},
  morestring=[b]"
}
\newtheorem{theorem}{Theorem}[section]

\newtheorem{definition}[theorem]{Definition}

\hypersetup{
  colorlinks=true,
  linkcolor=blue!70!black,
  citecolor=blue!70!black,
  urlcolor=blue!70!black
}

\usepackage{seqsplit}
\newcommand{\isaf}[1]{\texttt{#1}}
\newcommand{\hexid}[1]{\texttt{\seqsplit{#1}}}
\newcommand{\Applied}{\mathsf{Applied}}
\newcommand{\Rejected}{\mathsf{Rejected}}
\newcommand{\Operational}{\mathsf{OperationalFailure}}
\newcommand{\State}{\mathcal{S}}
\newcommand{\Obs}{\mathcal{O}}

\title{Mechanizing Typed Regulatory Actions for Security Tokens:\\Semantics, Falsification, and Bounded EVM Evidence}

\author{
  Jinwook Kim\textsuperscript{1,2}~\href{https://orcid.org/0009-0004-4993-8005}{\mbox{\scalerel*{\begin{tikzpicture}[yscale=-1,transform shape]\pic{orcidlogo};\end{tikzpicture}}{|}}} \\
  (for the Oraclizer Core Team) \\
  \textsuperscript{1}Oraclizer Labs, Delaware, USA \\
  \textsuperscript{2}Oraclizer Labs Korea, Seoul, Korea \\
  \texttt{jay@oraclizer.io}
}

\date{}
\begin{document}
\maketitle

\begin{abstract}
Security-token standards expose privileged controls without identifying the legal effect executed or the evidence and reversal obligations it carries. We formalize in Isabelle/HOL a reference execution semantics for the six ERC-8319 meanings: FREEZE, SEIZE, CONFISCATE, LIQUIDATE, RESTRICT, and RECOVER. It distinguishes applied, rejected, and operational-failure outcomes and mechanizes action-specific reversals, replay and epoch rules, complete frames, case-local terminality, and final receipts; the session builds without unproved placeholders or additional axioms. An indistinguishability theorem shows that bound kernel inputs cannot establish external facts about title, settlement, or entitlement. Constructive witnesses and direct mutations establish reachability and sensitivity for the declared fault set. For a successor ERC-TRUST Solidity/EVM candidate, we report separately scoped Foundry, Certora, Kontrol/KEVM, mutation, deterministic-build, runtime-identity, and independent-reproduction evidence. All twelve named evidence lanes pass with none pending, while the 74-row obligation ledger remains conditional: 70 rows are closed, two runtime-link rows remain successor obligations, and two are inapplicable. The Native runtime is bound separately from an ERC-3643 interoperability reference that explicitly reports Partial and \texttt{full=false}, not Verified Full. These results do not establish complete Isabelle-to-Solidity-to-EVM refinement, compiler correctness, audit completion, production readiness, deployment verification, or external legal truth. They provide a machine-checked domain semantics and an explicit map of proved, bounded, assumed, and open results.
\end{abstract}


\section{Introduction}
\label{sec:introduction}

Security-token contracts already contain powerful control mechanisms. ERC-3643 distinguishes forced transfer, recovery, freezing, unfreezing, pausing, and role-governed compliance~\cite{erc3643}; ERC-7943 defines a neutral forced-transfer and frozen-token interface~\cite{erc7943}; ERC-1450 gives a registered transfer agent exclusive execution authority and defines a request lifecycle with rejection reasons, terminal states, and one-time request identifiers~\cite{erc1450}. The open problem addressed here is therefore not the absence of privileged mechanics. It is that the on-chain observations of these standards do not identify which legally distinct effect a privileged movement executed: the same balance movement may serve custody, confiscation, recovery, or liquidation while carrying different evidence, reversal, terminality, and receipt obligations. Proposed ERC-8319, a companion proposal co-authored by the present author~\cite{erc8319}, names those six effects; this paper supplies the execution semantics and the evidence boundary for them.

The distinction matters because a successful call can be technically correct while its external premise is false. A contract may faithfully move a balance to a declared custodian without proving legal title, consume an entitlement commitment without proving rightful ownership, or bind a settlement reference without proving that a sale occurred. Conversely, a normal policy denial and an unavailable or malformed dependency may both revert while requiring different audit and retry treatment. A verification claim that collapses these cases can be stronger than its evidence.

This paper asks four questions:

\begin{enumerate}[label=\textbf{RQ\arabic*:},leftmargin=2.8em]
  \item How can the six ERC-8319 regulatory effects and their declared reversals be represented as a typed execution semantics without claiming external legal truth?
  \item Which replay, epoch, frame, terminality, outcome, and receipt properties can be machine-checked in the abstract model, and are their premises nonvacuous?
  \item Which relationships between that model and one Solidity/EVM candidate have current evidence, and which remain conditional or bounded?
  \item How should heterogeneous tool results and artifact identities be reported without adding them into a nonexistent end-to-end soundness theorem?
\end{enumerate}

\paragraph{Contributions.}
The work makes four bounded contributions. First, it instantiates the six-action vocabulary of proposed ERC-8319~\cite{erc8319} as a machine-checked reference execution semantics with separate reversal and three-way outcome domains. Second, it proves a domain theorem suite covering state frames, replay and stale epochs, case-local terminality, final receipts, failure stutter, and an indistinguishability boundary for external truth. Third, it supplies constructive witnesses evaluated from an executable manifest covering all six actions under all three outcome scenarios, together with mutation-based falsification for the declared semantic distinctions. Fourth, it presents a correspondence case study for one ERC-TRUST candidate,\footnote{ERC-TRUST is the working name of a pre-ERC draft specification. No ERC number has been assigned, no ethereum/ERCs pull request has been opened for it, and no editorial status is implied by the name.} separating model theorems, a unique and functional current-state abstraction relation, conditional profile-scoped refinement corollaries, Solidity rules and tests, selected EVM seams (narrow, high-risk correspondence points such as ABI decoding, external calls, and revert paths), artifact identity, and the absent deployment boundary.

The case study also contributes two disciplines that are reusable outside this standard: an evidence stratification in which no combination of tool passes may imply a stronger cross-layer theorem, and an artifact-identity regime that separates semantic runtime drift from packaging drift and rejects expected-hash overwrites. A stable-identifier objection register accompanies the paper so that any disposition can be overturned by a stronger standard, theorem, trace, or counterexample.

The paper does not introduce regulated-token control mechanisms, generic typed outcomes, nonce-based replay protection, smart-contract formal verification, mutation testing, or software provenance. It also does not establish a complete Isabelle-to-Solidity-to-EVM refinement theorem. The closest cross-layer comparators verify EVM bytecode in Isabelle, provide executable EVM semantics, embed Solidity deeply in Isabelle/HOL, or verify compilation to EVM~\cite{amani2018,kevm2018,isabellesolidity2025,deepsea2024}. The present result is a domain-specific semantics and evidence-boundary study.

\paragraph{Reading paths.}
Implementers of the standard can start with the overview and evidence sections (\cref{sec:overview,sec:implementation,sec:evidence}) and the correspondence and command appendices (Appendices~\ref{app:correspondence} and~\ref{app:commands}). Standards reviewers can start with the objection analysis (\cref{sec:objections}, Appendix~\ref{app:objections}). Formal-methods readers can start with the semantics and mechanization sections (\cref{sec:semantics,sec:mechanization,sec:countermodels}) and the theorem and assumption appendices (Appendices~\ref{app:theorems} and~\ref{app:assumptions}). Practitioners in regulated assets can read the overview (\cref{sec:overview}), the boundary column of \cref{tab:actions}, and the external and deployment limitations in \cref{sec:limits}. Appendix~\ref{app:terminology} defines the recurring technical vocabulary.

\section{Regulatory Execution Problem, System Model, and Non-Goals}
\label{sec:model}

\subsection{Mechanism, meaning, and assurance}

We separate three questions that are often merged in security-token discussions. A \emph{mechanism} is a callable transition such as a freeze or forced transfer. A \emph{meaning} identifies which regulatory effect that transition is intended to realize and which action-specific record it must preserve. An \emph{assurance statement} names the exact model, implementation, tool, and environmental assumptions supporting a claim. ERC-7943 deliberately chooses mechanism-neutral names; ERC-8319 supplies legal-effect vocabulary; ERC-TRUST is studied here as an execution and conformance layer between them.

The mechanized action vocabulary is bound to the ERC-8319 draft text as published in ethereum/ERCs pull request 1848 at commit \hexid{a7267cb3ce81e6e50fe6ef00b98ec98bb76264f8} (retrieved 2026-08-29). Editorial revisions of ERC-8319 after that commit create new conformance obligations for future versions of this artifact and are not retroactively covered.

\subsection{System model}

The abstract system contains a persistent state $s\in\State$, a typed command $c\in\mathsf{Command}$, an authorization and policy context $a\in\mathsf{Auth}$, an external observation $e\in\mathsf{External}$, an outcome $o\in\mathsf{Outcome}$, and a declared observation $\omega\in\Obs$. Commands are forward actions or reversals. A forward action belongs to
\[
  \mathsf{Action}=\{\mathsf{FREEZE},\mathsf{SEIZE},\mathsf{CONFISCATE},
  \mathsf{LIQUIDATE},\mathsf{RESTRICT},\mathsf{RECOVER}\},
\]
and the declared reversal domain contains $\mathsf{UNFREEZE}$, $\mathsf{RELEASE}$, and $\mathsf{UNRESTRICT}$. The kernel evaluates the command and its bindings and yields one of $\Applied$ (written \isaf{AppliedOnChain} in the artifact), $\Rejected$, or $\Operational$.

The concrete case study uses an immutable Solidity state owner, read-only fail-closed dependencies, a same-transaction exact-use ticket for sensitive ERC-7943 selectors, and an optional ERC-3643 Partial adapter profile. The reference candidate is not a claim that every conforming system must use this architecture.

\subsection{Threat model}

The model permits malformed commands, replay attempts, stale epochs, denied policy observations, dependency reverts, malformed return data, selector bypass attempts, duplicate reversals, out-of-order restoration, and inconsistent artifact identities. It also considers a dishonest external world that supplies a syntactically valid but substantively false attestation. The kernel is expected to fail closed on malformed or unavailable declared inputs and to preserve its state on rejected or operational outcomes.

The model does not assume that a regulatory authority is legitimate, that keys are uncompromised, that an oracle reports truth, that a settlement occurred, or that a legal entitlement belongs to the named person. These are distinct governance and external-world assumptions.

\subsection{Observation alphabet and non-goals}

Failure stutter, the requirement that a rejected or operationally failed step leaves state exactly unchanged, is declared over persistent kernel state, declared downstream state, committed logs, and return/revert classification. It does not cover gas consumption, timing, mempool observation, miner or validator ordering, or every side channel. The paper further excludes compiler correctness, arbitrary proxy and migration systems, arbitrary ERC-3643 deployments, live deployment identity, legal advice, audit completion, and production readiness.

\section{ERC-TRUST Overview}
\label{sec:overview}

\begin{figure}[ht]
  \centering
  \resizebox{\textwidth}{!}{
\begin{tikzpicture}[
  box/.style={rectangle, rounded corners=2pt, draw=black, fill=gray!4,
    minimum width=2.7cm, minimum height=0.85cm, align=center, font=\small},
  applied/.style={box, draw=green!45!black, fill=green!7},
  reject/.style={box, draw=orange!70!black, fill=orange!8},
  failure/.style={box, draw=red!60!black, fill=red!6},
  ext/.style={rectangle, rounded corners=2pt, draw=black!55, fill=white,
    dashed, minimum width=3.2cm, minimum height=0.75cm, align=center, font=\footnotesize},
  arr/.style={-{Stealth[length=5pt]}, thick},
  node distance=0.75cm and 0.9cm
]
  \node[box] (command) {Typed command\\action or reversal};
  \node[box, right=of command] (assess) {Canonical assessment\\binding, authority, evidence};
  \node[applied, right=1.05cm of assess, yshift=1.1cm] (applied) {AppliedOnChain\\state effect + final receipt};
  \node[reject, right=1.05cm of assess] (rejected) {Rejected\\policy completed, denied};
  \node[failure, right=1.05cm of assess, yshift=-1.1cm] (failure) {OperationalFailure\\dependency unavailable or malformed};
  \draw[arr] (command) -- (assess);
  \draw[arr] (assess.east) -- ++(0.45,0) |- (applied.west);
  \draw[arr] (assess) -- (rejected);
  \draw[arr] (assess.east) -- ++(0.45,0) |- (failure.west);
  \node[ext, below=1.75cm of command, xshift=1.35cm] (world) {External legal and factual world\\not determined by the kernel};
  \draw[arr, dashed, black!55] (world.north) -- node[left, font=\scriptsize, align=right]
    {bound input only} (assess.south);
\end{tikzpicture}}
  \caption{Typed regulatory execution. The three outcomes have different meanings even when two of them revert. External truth is deliberately outside the kernel observation.}
  \label{fig:semantics}
\end{figure}
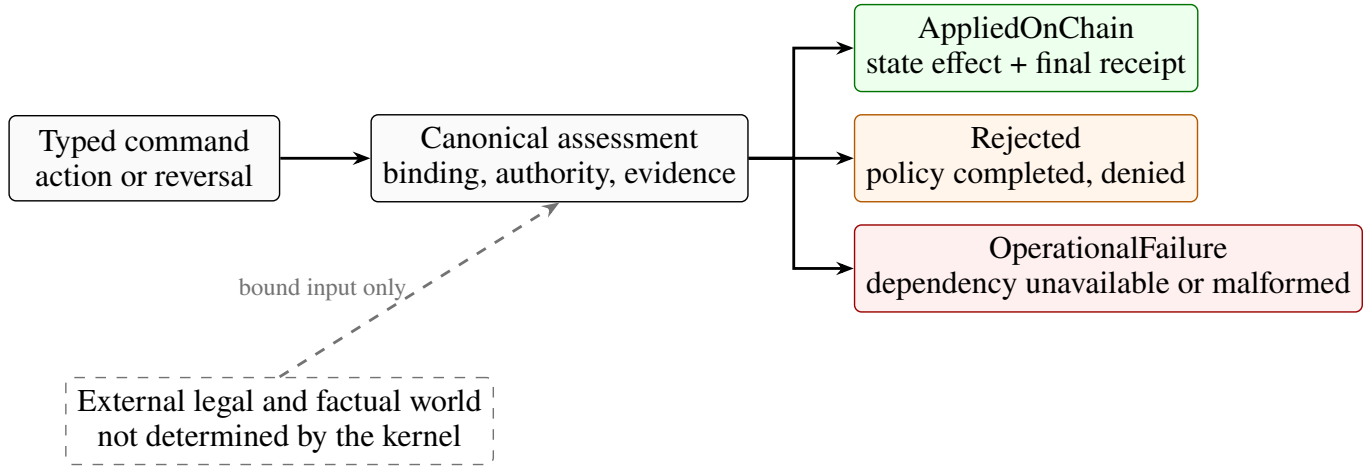

\Cref{fig:semantics} summarizes the execution pipeline: a typed command undergoes one canonical assessment and yields exactly one of three outcomes, while external legal and factual truth stays outside the kernel boundary.

\subsection{Actions and reversals}

\Cref{tab:actions} lists each action's declared reversal, its declared on-chain effect, and the external fact the kernel deliberately does not establish. A \emph{case} is an identifier for one regulatory proceeding, such as one seizure or liquidation matter; terminality is declared per case, never for the whole token.

\begin{table}[ht]
\centering
\caption{Declared action effects, reversals, and external-truth boundaries.}
\label{tab:actions}
\small
\begin{tabular}{@{}P{2.05cm}P{2.1cm}P{4.2cm}P{4.4cm}@{}}
\toprule
Action & Declared reversal & Declared on-chain effect & Not established by the kernel \\
\midrule
FREEZE & UNFREEZE & Strictly raise the absolute frozen target; an explicit reversal restores the referenced prior target & Legitimacy of the freezing order \\
SEIZE & RELEASE & Transfer into declared custody and record prior holder and encumbrance & Legal or beneficial title \\
CONFISCATE & none (terminal for the case) & Terminal privileged disposition for the case & Lawfulness or final adjudication \\
LIQUIDATE & none & Bind token disposition to settlement and proceeds commitments & Actual sale, proceeds receipt, or debt discharge \\
RESTRICT & UNRESTRICT & Set an ordinary-transfer restriction; the reversal restores the prior flag & Substantive validity of the restriction policy \\
RECOVER & none & Consume an entitlement commitment and credit a declared destination once & Rightful ownership or identity truth \\
\bottomrule
\end{tabular}
\end{table}

Reversal is not modeled as a second untyped balance operation. A reversal refers to an original action and its recorded prior state; a stale, duplicate, or out-of-order attempt is rejected with full persistent-state stutter, and an applied restoration produces a terminal record for that restoration attempt. CONFISCATE, LIQUIDATE, and RECOVER do not acquire an implicit reversal merely because another transfer could later be executed.

\subsection{Three outcomes}

$\Applied$ means that the declared on-chain transition and canonical receipt committed. $\Rejected$ means the required checks completed and denied the command. $\Operational$ means a required dependency was unavailable, malformed, stale, inconsistent, or otherwise unable to return a canonical assessment. The contribution is the regulatory-action-specific state, receipt, and audit semantics attached to the partition, not the general idea of typed status classes.

\subsection{Reference architecture}

\begin{figure}[ht]
  \centering
  \resizebox{0.96\textwidth}{!}{
\begin{tikzpicture}[
  core/.style={rectangle, rounded corners=2pt, draw=blue!65!black, fill=blue!6,
    minimum width=3.0cm, minimum height=1.0cm, align=center, font=\small},
  box/.style={rectangle, rounded corners=2pt, draw=black, fill=gray!4,
    minimum width=2.7cm, minimum height=0.85cm, align=center, font=\small},
  ext/.style={rectangle, rounded corners=2pt, draw=black!55, fill=white, dashed,
    minimum width=2.8cm, minimum height=0.8cm, align=center, font=\footnotesize},
  arr/.style={-{Stealth[length=5pt]}, thick},
  node distance=0.75cm and 0.95cm
]
  \node[ext] (authority) {Authority and policy world\\legitimacy not proved};
  \node[ext, below=of authority] (providers) {Identity, settlement, entitlement\\truth not proved};
  \node[core, right=1.1cm of authority, yshift=-0.8cm] (kernel)
    {Immutable minimal kernel\\typed execution + fail-closed checks};
  \node[box, right=of kernel] (state) {Single state owner\\effect record + receipt};
  \node[box, above=of state] (ticket) {ERC-7943 exact-use\\same-transaction ticket};
  \node[box, below=of state] (profile) {ERC-3643 Partial profile\\declared-entry adapter checks};
  \draw[arr, dashed, black!55] (authority.east) -- (kernel.north west);
  \draw[arr, dashed, black!55] (providers.east) -- (kernel.south west);
  \draw[arr] (kernel) -- (state);
  \draw[arr] (kernel.north east) |- (ticket.west);
  \draw[arr] (kernel.south east) |- (profile.west);
\end{tikzpicture}}
  \caption{The reference architecture. Bound external inputs are consumed by an immutable kernel and a single regulatory-state owner. The dashed boundary marks what the kernel cannot know from its bound inputs, a proved limit (\cref{thm:external}), not merely an implementation disclaimer.}
  \label{fig:architecture}
\end{figure}
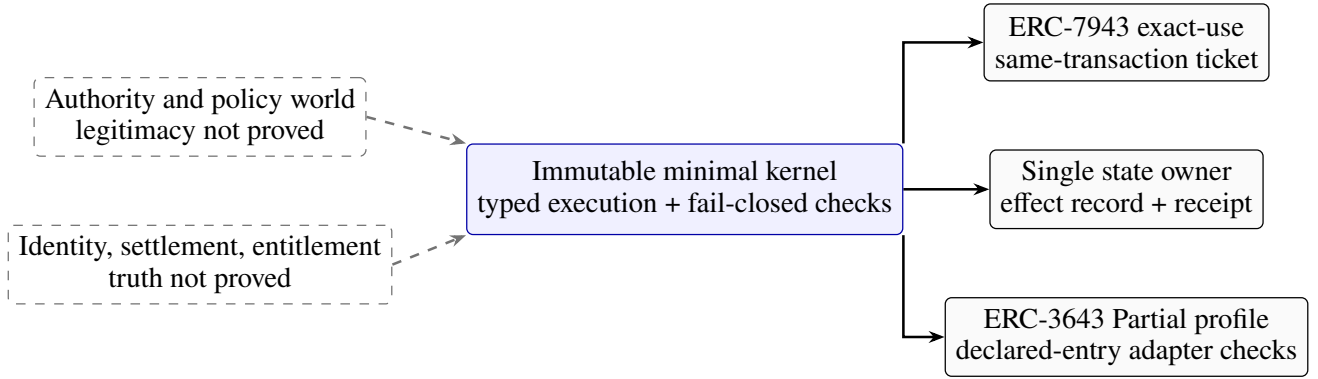

\Cref{fig:architecture} shows the reference topology. The architecture keeps regulatory state and receipts in one owner. Read-only dependency modules are bound by address, runtime code, schema, configuration, epoch, and command echo. Sensitive ERC-7943 calls require an exact-use ticket bound to caller, selector, calldata, policy binding, epochs, and command identifier, then consumed in the same transaction. ERC-3643 support is an optional Partial interoperability profile. Its adapter checks declared import entries, fail-closed touch points, actual restriction post-state, and actual restriction observations, but it does not establish manifest completeness or same-transaction enforcement for ordinary inbound transfers. It therefore reports \texttt{profileKind=PARTIAL} and \texttt{full=false}. A future Verified Full profile requires a fresh atomic deployment, a complete initial-state gate, and a same-transaction transfer or Compliance hook.

The reference candidate deliberately avoids upgradeable proxy and Diamond patterns. Delegatecall indirection, storage-slot aliasing, and admin-controlled implementation substitution would multiply the state-correspondence and runtime-identity obligations that the present evidence discharges, and would break the exact-runtime premises used by the retrieve relation. This is a proof-cost decision for the reference candidate, not a claim that a conforming implementation must be non-upgradeable.

\section{Typed Regulatory Execution Semantics}
\label{sec:semantics}

\subsection{State, commands, and observations}

The state record includes balances, allowances, frozen targets, restriction flags, case state, custody and encumbrance records, action and reversal histories, consumed nonces and entitlements, authority and policy epochs, configuration bindings, receipts, and auxiliary state. An applied step commits an ordered sequence of observations, so receipt finality is well defined over that order. Each operation declares a write set. This makes a frame theorem falsifiable: fields outside the declared set must be preserved. The declared write sets are not free to inflate: a separate exact-effects theorem pins each applied transition to its declared effect fields, and a frame-removal mutation must break the complete-frame consumer (\cref{tab:mutations}).

A command binds domain, implementation, chain, action or reversal kind, source, destination, amount, case, authority and policy epochs, nonce, validity interval, provenance commitment, and action-specific evidence. The exact concrete encoding belongs to the implementation relation, not the abstract definition.

\begin{definition}[Canonical execution]
For state $s$, command $c$, authorization context $a$, and external observation $e$, canonical execution is a total function
\[
  \mathsf{exec}:\State\times\mathsf{Command}\times\mathsf{Auth}\times\mathsf{External}
  \rightarrow \State\times\mathsf{Outcome}\times\Obs.
\]
It either applies the typed effect and writes one final canonical receipt, or returns a rejected or operational outcome with declared persistent-state stutter.
\end{definition}

\begin{theorem}[Outcome classification and stutter]
Within the declared abstract domain, canonical execution assigns exactly one of $\Applied$, $\Rejected$, or $\Operational$ to every step. If it returns $\Rejected$ or $\Operational$, its post-state equals its pre-state over the complete persistent state record.
\end{theorem}

Disjointness of the outcome constructors holds by construction; the content of the theorem is the total classification and the machine-checked full-state stutter for canonical rejection and operational failure, together with their entrypoint and transaction-level refinements inside the model. The exact Isabelle names are listed in Appendix~\ref{app:theorems}. The theorem does not assert equal gas or timing.

\begin{theorem}[Entrypoint convergence]
Within the declared model, every typed entrypoint normalizes to the same canonical executor, and the untyped legacy entrypoint fails closed.
\end{theorem}

On an applied result, the mechanized post-state fixes the exact regulatory mode, balance and custody effects, case record, authorization and nonce consumption, pre/post observation commitments, and final canonical receipt. On rejected or operational failure, the complete persistent state stutters, and the success theorem preserves every unrelated field named by the full frame.

\subsection{Replay, epochs, and frames}

Replay protection combines a domain-bound command identifier with an authority-epoch and nonce tuple. Successful execution consumes the identifier and nonce. Authority rotation and policy rebinding invalidate stale authorizations. ERC-3009 already supplies validity windows, cancellation, and single-consumption authorization~\cite{erc3009}; the specific contribution here is their integration with typed regulatory effects, case dynamics, epochs, and action-specific receipts.

\begin{theorem}[Frame and final receipt]
For an applied canonical action, every persistent field outside its declared write set is preserved, and the applied receipt is the final persistent receipt observation of the step. Rejected and operational outcomes commit no receipt.
\end{theorem}

\begin{theorem}[Reversal admissibility]
A reversal command refers to a specific original action and its recorded prior state. A stale, duplicate, or out-of-order reversal is rejected with full persistent-state stutter; an applied reversal restores the referenced prior state and commits a terminal record for that restoration attempt.
\end{theorem}

\subsection{Case-local terminality}

Terminality is indexed by case rather than global token state: a machine-checked theorem scopes terminality to the affected case, so unrelated cases and assets proceed after one case becomes terminal. A deliberately conflated global-terminal mutation is distinguished by the current-profile state theorem.

\subsection{External-truth claim boundary}

\begin{theorem}[External-truth non-identifiability]
\label{thm:external}
There exist two concrete external worlds that disagree about retained title, whether a sale occurred, whether debt was discharged, and whether the recipient is rightful, yet for every kernel state and entry they induce the same core observation. The kernel therefore cannot establish any of those external facts from its bound inputs alone.
\end{theorem}

The mechanized statement is universally quantified over kernel states and entries, with the two disagreeing worlds given as explicit record values:

\begin{lstlisting}
theorem same_core_input_cannot_distinguish_external_legal_truth:
  "core_observation world_with_all_external_truths st entry =
   core_observation world_with_no_external_truths st entry"
  by (simp add: core_observation_def)

theorem external_worlds_really_disagree:
  "world_title_retained world_with_all_external_truths \<noteq>
     world_title_retained world_with_no_external_truths \<and>
   world_sale_occurred world_with_all_external_truths \<noteq>
     world_sale_occurred world_with_no_external_truths \<and>
   world_debt_discharged world_with_all_external_truths \<noteq>
     world_debt_discharged world_with_no_external_truths \<and>
   world_recipient_rightful world_with_all_external_truths \<noteq>
     world_recipient_rightful world_with_no_external_truths"
  by (simp add: world_with_all_external_truths_def
      world_with_no_external_truths_def)
\end{lstlisting}

Here \isaf{core\_observation} is the artifact's projection of the kernel state and entry that canonical execution may depend on; the external-world fields are deliberately excluded from it. This is a positive limitation theorem: it prevents an implementation or paper from silently upgrading a checked commitment into a legal or factual truth claim.

Three actions make the boundary concrete in the mechanization, each as an assume-guarantee theorem placed next to its nonclaim. For SEIZE, the model guarantees custody accounting, the declared prior-holder reference, and encumbrance preservation, while actual legal or beneficial title remains a nonclaim. For LIQUIDATE, it binds the token disposition to settlement and proceeds commitments without asserting that a sale occurred, proceeds were received, or debt was discharged. For RECOVER, it guarantees the declared provider destination, the entitlement commitment, and its one-time consumption, while rightful ownership remains a nonclaim. A companion theorem fixes the exact inventory of claims the model is permitted to make, so a silently added claim fails the inventory check.

\section{Isabelle/HOL Mechanization and ERC-TRUST Instantiation}
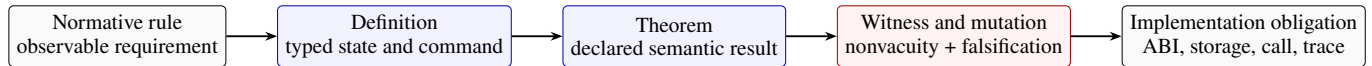
\label{sec:mechanization}

\begin{figure}[ht]
  \centering
  \resizebox{\textwidth}{!}{
\begin{tikzpicture}[
  stage/.style={rectangle, rounded corners=2pt, draw=black, fill=gray!4,
    minimum width=2.45cm, minimum height=0.9cm, align=center, font=\small},
  mech/.style={stage, draw=blue!60!black, fill=blue!5},
  neg/.style={stage, draw=red!55!black, fill=red!5},
  arr/.style={-{Stealth[length=5pt]}, thick},
  node distance=0.75cm
]
  \node[stage] (norm) {Normative rule\\observable requirement};
  \node[mech, right=of norm] (def) {Definition\\typed state and command};
  \node[mech, right=of def] (thm) {Theorem\\declared semantic result};
  \node[neg, right=of thm] (fal) {Witness and mutation\\nonvacuity + falsification};
  \node[stage, right=of fal] (impl) {Implementation obligation\\ABI, storage, call, trace};
  \draw[arr] (norm) -- (def);
  \draw[arr] (def) -- (thm);
  \draw[arr] (thm) -- (fal);
  \draw[arr] (fal) -- (impl);
  \node[font=\footnotesize, align=center, below=0.72cm of thm, xshift=1.6cm] (note)
    {Each edge has its own evidence class. A missing edge cannot be\\reconstructed by adding PASS results from adjacent tools.};
\end{tikzpicture}}
  \caption{From normative prose to a falsifiable implementation obligation. The proof and implementation relations are separate edges: proving the model does not by itself discharge the implementation obligation.}
  \label{fig:normative-trace}
\end{figure}

\Cref{fig:normative-trace} traces the pipeline this section mechanizes: a normative sentence becomes a typed definition, a theorem, a witness-and-mutation obligation, and finally an implementation obligation, and each edge carries its own evidence class.

The candidate's Isabelle2025-2~\cite{nipkow2002} session contains 22 top-level theory files. The session builds cleanly, and the proof-source audit records 412 explicit roots, 413 qualified facts, zero banned source forms (such as \isaf{sorry} or \isaf{oops}), and zero proof oracles beyond the standard HOL foundation. These counts describe the source inventory; they are not a count of independent contributions. The load-bearing theory groups are summarized in \cref{tab:theory-groups} and listed more precisely in Appendix~\ref{app:theorems}.

The mechanization keeps the six legal-effect kinds and the seven foundation transition labels as deliberately separate inventories:

\begin{lstlisting}
definition all_transition_labels :: "reg_action list" where
  "all_transition_labels =
    [FREEZE, SEIZE, CONFISCATE, RESTRICT, UNFREEZE, UNRESTRICT, RELEASE]"

definition all_rcp_actions :: "legal_action_kind list" where
  "all_rcp_actions =
    [Legal_Freeze, Legal_Seize, Legal_Confiscate, Legal_Restrict,
     Legal_Recover, Legal_Liquidate]"
\end{lstlisting}

RECOVER and LIQUIDATE are transfer-layer legal actions with no dedicated foundation label, and RELEASE is a reversal label with no forward legal meaning (\isaf{rcp\_action\_of\_forward\_label RELEASE = None}); a named theorem keeps the two inventories from being conflated.

\begin{table}[ht]
\centering
\caption{Mechanization groups and representative results.}
\label{tab:theory-groups}
\small
\begin{tabular}{@{}P{3.2cm}P{4.0cm}P{4.15cm}@{}}
\toprule
Group & Representative result & Boundary \\
\midrule
Action mapping & Six actions are distinct from transition labels and reversals & Imports ERC-8319 vocabulary; does not invent it \\
Canonical semantics & Outcome separation, failure stutter, frames, final receipt & Declared abstract state and observation alphabet \\
Simulation and witnesses & Reachable action/outcome and reversal cases & Nonvacuity, not model adequacy by itself \\
Compatibility & ERC-7943 typed binding and explicit ERC-3643 subset & Abstract mapping, not arbitrary deployment proof \\
Claim boundary & Same input cannot distinguish disagreeing external worlds & No claim about which external world is true \\
Transaction relation & Success, failure, malformed input, and receipt consequences & Modeled transaction relation, not compiler correctness \\
Conditional composition & Typed consequences under a runtime-link premise & Runtime link is an assumption, not a discharged central theorem \\
Current profile & Package-conditioned row corollaries and distinguishing negatives & Exact candidate and receipt premises \\
\bottomrule
\end{tabular}
\end{table}

In the mechanized compatibility model, a strict ERC-7943 frozen-amount increase converges to typed FREEZE. An equal or decreasing raw forward call is rejected; a decrease requires a separately authorized UNFREEZE reversal. A forced transfer requires a typed binding, and an unbound forced transfer fails closed. A further theorem shows that the compatibility mapping does not expand the claim boundary. These results hold in the abstract compatibility model; they are not proofs about arbitrary third-party token deployments.

\subsection{The conditional theorem named end-to-end refinement}

The source contains a theorem named \isaf{end\_to\_end\_refinement}. Its name must not be read as a completion claim. It is inside locale \isaf{pinned\_runtime\_refinement}, whose \isaf{runtime\_link} assumption already states that a runtime execution satisfies the abstract transaction relation. The locale header makes the assumption explicit:

\begin{lstlisting}
locale pinned_runtime_refinement =
  fixes manifest :: trust_runtime_manifest
    and bridge :: trust_transaction_bridge
    and certificate :: trust_runtime_certificate
    and assumptions :: trust_external_assumptions
    and runtime_execution :: "trust_transaction_execution \<Rightarrow> bool"
    and runtime_abstraction :: "trust_transaction_execution \<Rightarrow> trust_transaction_abstraction"
  assumes certificate_complete: "runtime_certificate_complete certificate"
      and external_assumptions: "external_assumptions_complete assumptions"
      and runtime_link:
        "runtime_execution execution \<Longrightarrow>
         alpha_transaction manifest bridge execution (runtime_abstraction execution)"
\end{lstlisting} The theorem repeats that premise and exposes a useful typed consequence interface. The missing work is a machine-checked construction of that premise from the compiler, runtime semantics, and concrete execution. The development also records compiler correctness as an explicitly named nonclaim (\isaf{compiler\_correctness\_remains\_a\_nonclaim}) rather than an implied consequence. The theorem name \isaf{end\_to\_end\_refinement} is retained for artifact-identity stability; its correct reading is fixed here and in Appendix~\ref{app:theorems}.

\subsection{Foundation and instantiation}

The regulatory model uses the state-machine foundations of Cross-Domain State Preservation~\cite{cdsp2026} and its independently released \emph{Regulatory Action Composition} (RAC) session~\cite{racartifact}. RAC studies outcome-sensitive sequential composition over the five-state, seven-transition reference machine: it separates applied actions, legal rejections, and operational failures, and establishes action-order effects, terminal behavior, observable traces, and finite transformation normal forms. These are inherited results. The present development adds the typed six-action execution model and action-specific reversals, then relates their state and receipt consequences to separately scoped implementation evidence; it does not re-count RAC's classifications as new contributions.

The product execution semantics imports RAC under the foundation session's qualified name. RAC is published as a separate child session in the \href{https://github.com/Oraclizer/formal-verification/tree/db8e0802e55f0229cf7bc9e5e6cfbf40681adbba/Regulatory_Action_Composition}{\texttt{Oraclizer/formal-verification} repository}. The candidate's public dependency lock pins that revision and constructs a temporary compatibility session for the qualified import. The mapped RAC theorem statements and bodies are preserved; the title comment and qualified import path account for the recorded source differences. This build arrangement neither merges the ownership of the two developments nor establishes model-to-code refinement.

Because the historical dependency object became unreachable at its public remote, the candidate records file-by-file succession for fourteen foundation inputs. Negative probes cover theorem-body mutation and wrong-import and wrong-commit substitutions, separating proof-content failure from supply-chain identity failure. The successor session also imports a generated runtime bridge that fixes the three compiled runtime identities and an obligation-ledger theory that names every abstract condition consumed by the current implementation map. The ledger has 74 rows: 70 closed, two successor-mandatory runtime-link rows, two not applicable, and no current-mandatory row. These records fix the formal inputs consumed by the current Isabelle build and expose the remaining central link; they do not show that the Solidity candidate refines every foundation theorem.

\section{Concrete Nonvacuity and Direct Falsification}
\label{sec:countermodels}

Falsification evidence throughout means a deliberately mutated system in which a removed condition makes its consumer theorem or check fail, showing that the condition is load-bearing.

\subsection{Constructive reachability}

The executable manifest is an evaluated artifact rather than a copied outcome table: its 18 distinct inputs are exactly the Cartesian product of six actions and three outcome scenarios, and each row's outcome, target state, required observable, and write set is computed from the executable manifest semantics. Isabelle supplies reachable applied, authorization-denial, and policy-unavailable witnesses for every action; separate reachable witnesses cover all three reversals, ordinary-transfer success and denial, and alternate-overlay compatibility paths. These witnesses prevent a theorem from passing merely because its premise is uninhabited. They do not by themselves establish model adequacy.

\subsection{Direct falsification}

The abstract negative campaign removes or changes load-bearing conditions such as action/receipt identity, epoch invalidation, outcome separation, frame preservation, and case-local terminality; \cref{tab:mutations} names the consumer that each removed distinction must break. A direct mutation must change the consumer theorem or verification result it is intended to test. At the pinned candidate, the abstract campaign comprises 15 direct mutations, and each is detected by the consumer it targets. This is narrower than mutation adequacy in general; Solidity-specific mutation testing predates this work~\cite{sumo2021}.

\begin{table}[ht]
\centering
\caption{Representative falsification obligations.}
\label{tab:mutations}
\small
\begin{tabular}{@{}P{3.25cm}P{4.35cm}P{3.8cm}@{}}
\toprule
Removed distinction & Expected counterexample & Consumer \\
\midrule
Action/receipt identity & A receipt can be replayed or attributed to the wrong effect & Receipt and replay theorems \\
Policy or authority epoch & A stale authorization remains admissible & Lifecycle and stale-epoch theorems \\
Rejected/operational split & A dependency outage becomes an ordinary policy denial & Outcome partition and audit semantics \\
Frame condition & Unrelated state can change during an action & Complete-frame theorem \\
Case-local terminality & One terminal case blocks unrelated cases & Scoped-terminality theorem \\
Prior-state reversal link & Stale, duplicate, or out-of-order restoration succeeds & Reversal admissibility and restoration \\
\bottomrule
\end{tabular}
\end{table}

Concrete positive and negative paths add evidence across a different layer. They do not prove the abstract model correct; they make specific correspondence claims falsifiable.

\section{Relationship to the Solidity and EVM Implementation}
\label{sec:implementation}

\subsection{Abstraction relation}

The current retrieve (abstraction) relation maps concrete balances, frozen targets, restriction flags, custody records, case terminality, nonces, epochs, and receipts into an abstract state. Under its declared well-formedness and pinned-runtime premises, it is unique and functional, rejects runtime substitution, and projects nonces, freeze and restriction overlays, custody, and case terminality exactly. The abstraction is a function of the currently bound on-chain state and pinned configuration (dependency topology, endpoint identity, manifest, and storage footprint), not of any assumed execution history. The transaction relation then connects modeled concrete executions with applied, rejected, operational, malformed-input, and dependency-revert outcomes.

This narrows the model-to-code gap but remains below verified compilation. Amani et al. construct a sound bytecode logic in Isabelle/HOL~\cite{amani2018}; KEVM supplies executable EVM semantics and bytecode verification~\cite{kevm2018}; Isabelle/Solidity deeply embeds Solidity in Isabelle/HOL~\cite{isabellesolidity2025}; DeepSEA supplies a verified compiler and end-to-end case studies~\cite{deepsea2024}. The present candidate instead combines a hand-written Solidity implementation, conditional Isabelle relations, bounded Solidity verification, and selected compiled-EVM seams.

\subsection{ABI, storage, calls, and traces}

The correspondence map names the following obligations:

\begin{itemize}
  \item canonical ABI decoding and rejection of malformed length, enum, address, and fixed-width values;
  \item storage projection for balances, state overlays, custody, case terminality, nonces, epochs, and receipts;
  \item low-level read-only dependency call and return-data classification;
  \item same-transaction exact-use ticket creation, consumption, and deletion;
  \item action and reversal effects, rollback, and complete failure stutter;
  \item event order and final canonical receipt;
  \item creation/runtime bytecode, method identifiers, storage layout, and immutable-reference identity.
\end{itemize}

The fixed-width decoder result uses an explicit cross-kernel composition: KEVM proves conditional runtime reachability, Isabelle proves that the listed word bounds hold over the fixed 256-bit domain, and a hash-bound record states the K/Isabelle correspondence. No single kernel checks that last correspondence, so it remains in the trusted computing base.

\subsection{Open central obligation}

The open result is a proof object that constructs the runtime-link relation for every accepted execution in the declared profile from source, compiler or bytecode semantics, external-call assumptions, and trace observations. A deep embedding of Solidity in Isabelle/HOL~\cite{isabellesolidity2025} is one candidate substrate for constructing that relation; evaluating its coverage of the candidate's Solidity 0.8.36 feature set and storage model is part of the open obligation. Nevertheless, the current version includes conditional, profile-scoped refinement theorems: named package and row corollaries derive typed state, failure, and receipt consequences once their hash-bound qualification certificates and abstract-summary premises hold. We reserve \emph{complete end-to-end refinement} for a theorem that constructs the runtime link rather than assumes it.

\section{Layered Implementation Evidence}
\label{sec:evidence}

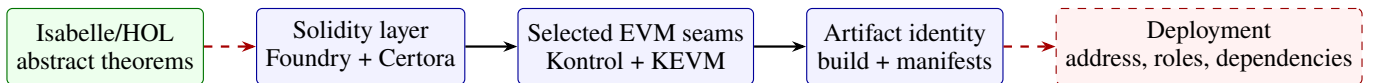
\begin{figure}[ht]
  \centering
  \resizebox{\textwidth}{!}{
\begin{tikzpicture}[
  layer/.style={rectangle, rounded corners=2pt, draw=black, fill=gray!4,
    minimum width=2.45cm, minimum height=0.95cm, align=center, font=\small},
  proved/.style={layer, draw=green!45!black, fill=green!7},
  bounded/.style={layer, draw=blue!60!black, fill=blue!5},
  open/.style={layer, draw=red!60!black, fill=red!5, dashed},
  arr/.style={-{Stealth[length=5pt]}, thick},
  gap/.style={-{Stealth[length=5pt]}, thick, dashed, red!65!black},
  node distance=0.65cm
]
  \node[proved] (isa) {Isabelle/HOL\\abstract theorems};
  \node[bounded, right=of isa] (sol) {Solidity layer\\Foundry + Certora};
  \node[bounded, right=of sol] (evm) {Selected EVM seams\\Kontrol + KEVM};
  \node[bounded, right=of evm] (id) {Artifact identity\\build + manifests};
  \node[open, right=of id] (dep) {Deployment\\address, roles, dependencies};
  \draw[gap] (isa) -- (sol);
  \draw[arr] (sol) -- (evm);
  \draw[arr] (evm) -- (id);
  \draw[gap] (id) -- (dep);
  \node[font=\footnotesize, align=center, below=0.72cm of evm] (note)
    {Green denotes machine-checked results in the declared model; blue denotes tool-specific evidence.\\Dashed red edges remain explicit obligations. Solid edges hand evidence to the next layer;\\they are not logical composition.};
\end{tikzpicture}}
  \caption{Evidence layers and open boundaries. Tool-specific PASS results do not create a stronger upward edge.}
  \label{fig:evidence-ladder}
\end{figure}

\Cref{fig:evidence-ladder} arranges the evidence classes and marks the two open edges: the model-to-Solidity relation and the deployment boundary.

\subsection{Current profile}

A publication profile is the declared set of implementation obligations that the candidate claims at release. Each obligation row names one checkable requirement; a reusable package groups the expensive runtime evidence that several rows share; and a qualification certificate is a hash-bound record that binds a row's evidence to the exact artifacts it covers. Qualification is a published, repository-defined pass criterion, not a third-party certification, audit, or regulatory approval. The successor release index names twelve evidence lanes: Foundry, mutation, Isabelle build, Isabelle/runtime binding, obligation ledger, Kontrol, Kontrol inputs, Certora, Certora inputs, independent reproduction, deterministic runtime, and runtime binding. All twelve pass and none is pending. The obligation ledger is a different projection: 70 of 74 rows are closed, two runtime-link rows remain successor obligations, and two are not applicable, so closure remains conditional. The preserved package projection separately reports 7/7 reusable packages, 49/49 Core obligations, 24/24 mandatory Supporting obligations, and 0/6 optional Verified Full obligations, with no partial credit. These denominators do not describe 73 independent whole-runtime proofs, and none implies the missing central theorem.

These fractions are not a progress score toward an unconditional theorem. They state that every row in a declared publication profile has a current qualification object. The central cross-layer theorem remains absent.

\subsection{Foundry and Certora}

At the pinned candidate, the implementation reports 93 Foundry~\cite{foundry} tests across seven suites, two fuzz properties with 256 runs each, and nine stateful invariants with 256 runs by 500 calls, for 1,152,000 invariant calls and zero invariant reverts. These are bounded tests, and the fuzz and invariant budgets are search budgets, not coverage claims. Certora~\cite{certoradocs} reports four of four named ERC-3643 Partial rules passing with advanced sanity. They establish that the descriptor is always Partial and never Full, that actual restriction post-state must match, and that subject- and role-side receipt observations read actual upstream restriction flags. The receipt binds the exact nine-file input root and current adapter runtime. These rules do not prove the Native endpoint, manifest completeness, ordinary-transfer hook coverage, every external dependency, or the whole runtime. A successful rule applies only to its exact harness, summaries, and prover assumptions.

\subsection{Kontrol and KEVM}

Kontrol~\cite{kontroldocs} reports four of four declared current-candidate proofs passing: raw sensitive selectors, operational-failure rollback, equal-or-decreasing FREEZE stutter, and liquidation delta with the final receipt log. Additional KEVM claims target ABI boundaries, dependency calls, receipt order, decoder guards, and package-to-row consequences. KEVM provides an executable EVM semantic foundation~\cite{kevm2018}, but using it on selected claims does not mean every runtime path was symbolically verified.

\subsection{Mutation and artifact identity}

At the Solidity level, a campaign of 121 declared implementation mutations reports all 121 killed and none surviving; this is detector evidence for the declared faults, not an audit. The two-layer runtime qualification checks three subjects, the Native token, the ERC-3643 Partial adapter, and its governor, across ABI, semantic storage layout, creation bytecode, runtime bytecode, method identifiers, and immutable-reference positions. Its verifier kills all 18 declared self-mutations and rejects stale receipts. An independent implementation written from the machine specification, generated prose and ABI, and conformance vectors reproduces 23 vectors and 401 assertions; four endpoint-wide or external-state items are explicitly not evaluable from that fixture. The release manifest binds UTF-8 text inputs after line-ending canonicalization, and deterministic builds bind source and compiler settings to creation and runtime bytecode. Exact-match artifact verification is established practice in the Ethereum ecosystem~\cite{sourcify}; its use here supports reproducibility rather than a new provenance theory.

\subsection{What the current evidence supports}

For a reader who builds the candidate at the stated commit, the current evidence supports exactly this. In the abstract model, rejected and operational outcomes provably leave every declared persistent field unchanged, and an applied action writes exactly its declared write set and one final receipt. The corresponding Solidity behaviors are exercised by the Foundry suite, the bounded fuzz and invariant runs, and the configured Certora Verification Language (CVL) rules under their stated assumptions. Selected compiled-EVM paths for ABI decoding, dependency classification, rollback, and receipt order carry symbolic evidence. The shipped bytecode is reproducibly bound to the stated source and compiler settings. None of this is an audit, a whole-program proof, or production readiness (\cref{sec:limits}).

\section{Artifact and Reproducibility}
\label{sec:artifact}

The paper is bound to the following exact artifact coordinates for the reviewed candidate. Repository merge status is an external release event and does not change the hash-bound evidence below.

\begin{center}
\small
\begin{tabular}{@{}ll@{}}
\toprule
Repository & \path{github.com/Oraclizer/erc-trust}~\cite{trustartifact} \\
Candidate & \isaf{0.2.0-candidate.1} (working label, no tag or release) \\
Reviewed public main & \hexid{a1cd93c8288b2a92655a9daf2e96d7f564cda614} \\
Git tree & \hexid{ebf2b4401bae23f37558e3a1a0953e40d507576e} \\
Release-manifest source root & \hexid{2b11f71daf04201e0ebc753820e3fba5875c9b2c65ae4d4ddd94a1e19b4cdf71} \\
Evidence input source root & \hexid{947afab7619b5c2d629da3e233c1cbe76fc6556a61112fbcc51b99c7e7235272} \\
Minimum replay entry & \path{scripts/replay-current-profile-release.ps1} \\
Replay verifier & \path{scripts/verify-current-profile-release-v3.mjs} \\
\bottomrule
\end{tabular}
\end{center}

The repository contains Isabelle theories, Solidity source and tests, Certora specifications, Kontrol/KEVM claims, mutation scripts, conformance vectors, schemas, runtime bindings, and replay scripts~\cite{trustartifact}.

Appendix~\ref{app:commands} lists the full command set. Model closure, model mutations, Foundry, Certora, Kontrol/KEVM, deterministic-build, runtime-binding, and release-identity checks remain separate commands because they consume different environments and establish different facts.

Reproduction of every command would show that the published artifact regenerates its declared results under the pinned environments. It would still not prove compiler correctness, tool implementation correctness, deployment identity, or external legal truth.

\section{Adversarial Analysis and Standardization Objections}
\label{sec:objections}

Each objection is stated in the strongest form we could construct, under stable identifiers intended to survive public review. The body discusses those with the broadest scope; Appendix~\ref{app:objections} lists all 22 with dispositions and residual limits.

\subsection{Existing standards already contain typed controls}

\textbf{OBJ-01.} ERC-3643 already separates recovery and freezing, ERC-1450 defines a request state machine, and ERC-7943 defines precise neutral mechanics. TRUST cannot claim that typed regulated-token controls were absent. Its narrower increment is a common semantics for the complete six-effect vocabulary where custody, confiscation, liquidation, and recovery may share a balance-movement mechanism but have different state, evidence, terminality, and receipt obligations.

\textbf{OBJ-02.} The contract cannot judge legal truth. We accept this objection as a design boundary and prove it in the model. Applied means that a declared transition committed under bound inputs; it is not a legal judgment.

\textbf{OBJ-03.} A six-action vocabulary may be too rigid across jurisdictions. The model separates effect vocabulary from jurisdiction policy. A jurisdiction that requires a new observable effect needs a new action or profile and a corresponding proof update; it should not silently overload an existing effect.

\subsection{Execution distinctions may duplicate established patterns}

\textbf{OBJ-04.} Action/reversal separation is not new in itself because freeze/unfreeze and pause/unpause precede this work. The domain-specific result is the combination of referenced prior state, stale and duplicate rejection, case-local terminality, and typed receipt identity.

\textbf{OBJ-05.} Expected rejection and exceptional failure also predate this work. The relevant result is their connection to regulatory audit/retry meaning and complete persistent-state stutter.

\textbf{OBJ-06 and OBJ-07.} Nonces and single consumption are established, and the exact-use ticket could be read as one more one-time authorization pattern. The candidate's narrower mechanism binds the legacy selector, caller, calldata, policy, epochs, and command into a same-transaction ticket and deletes it immediately; raw sensitive selectors remain closed.

\subsection{The proof may validate only its own model}

\textbf{OBJ-09.} The Isabelle-to-runtime gap is real. The paper exposes rather than hides it. A conditional theorem cannot discharge its own runtime-link assumption.

\textbf{OBJ-10.} Tool PASS results do not compose automatically. Each result remains in its own evidence class.

\textbf{OBJ-12.} Mutation sensitivity is not proof adequacy. It covers only the declared fault set.

\textbf{OBJ-13.} Nonvacuity witnesses may inhabit a wrong model. Cross-layer positive and negative evidence makes selected disagreements observable, while model adequacy remains open to specification review and counterexamples.

\subsection{Operational and governance limits}

\textbf{OBJ-15.} A pinned compiler may miscompile. The compiler remains trusted; bytecode evidence and provenance are reported separately.

\textbf{OBJ-16.} The verified candidate may differ from a deployment. This version makes no deployment claim.

\textbf{OBJ-17.} A malicious authority can execute a wrongful command through verified code. The model constrains scope, epoch, cancellation, state, and receipt; it does not establish legitimacy or key security.

\textbf{OBJ-20.} Runtime size remains a release constraint, but the successor is no longer near the EIP-170 ceiling. Two isolated clean builds agree on a 20,043-byte Native runtime, leaving 4,533 bytes; the ERC-3643 adapter is 19,480 bytes with 5,096 bytes left, and the governor is 2,787 bytes. Runtime binding establishes identity, not constructor execution or deployment. Compiler or feature drift can still consume the available margins and must rerun the size gate.

\textbf{OBJ-21.} ERC-8319 was proposed through ethereum/ERCs pull request 1848 and was not a canonically merged dependency at the pinned input commit. ERC-TRUST has no assigned ERC number, and no pull request for ERC-TRUST itself has been opened in ethereum/ERCs; an official submission with intended header \isaf{requires: 20, 165, 7943, 8319} is planned only after ERC-8319 is merged. These facts are stated as of this version and are updated in any future revision.

A public reviewer may overturn any initial disposition by producing a stronger standard, theorem, trace, or counterexample; new identifiers can be added without renumbering existing ones.

\section{Limitations, Trusted Computing Base, and Nonclaims}
\label{sec:limits}

\subsection{Trusted computing base}

The trusted computing base includes Isabelle2025-2 and its HOL libraries, the pinned K/KEVM/Kore stack and solver, Solidity 0.8.36, the Foundry 1.7.1 and Certora toolchains, Kontrol harnesses, manifest and hashing scripts, ABI and storage extractors, the explicit K/Isabelle word correspondence, operating-system and filesystem behavior relevant to replay, and human review of specification-to-code mappings. Some components provide proof kernels; others are ordinary programs whose correctness is not proved here.

\subsection{Model limitations}

The model abstracts gas, timing, mempool and sequencing behavior, denial of service, reentrancy beyond the declared architecture, and the full range of third-party dependency behavior. It fixes a reference action vocabulary and a current immutable architecture. Proxy, migration, arbitrary batch profiles, and broader ERC-3643 topologies require new conformance and evidence.

\subsection{Implementation limitations}

No complete machine-checked central refinement theorem connects all Isabelle states and traces to all Solidity/EVM executions. Selected EVM seams and bounded rules do not imply whole-program correctness. Two ledger rows remain successor obligations for that runtime link, and the six optional Verified Full obligations remain unclaimed. The current ERC-3643 profile is Partial: it does not establish a complete imported initial state or same-transaction enforcement for ordinary inbound transfers. The reference candidate is unaudited and not for production.

\subsection{External and deployment limitations}

No theorem proves authority legitimacy, policy correctness, identity, settlement, proceeds, title, entitlement, or rightful ownership. No deployed chain, address, constructor argument, role assignment, or live dependency is bound in this version. A future deployment manifest would establish identity for one deployment, not every deployment or its operational safety.

\section{Related Work and Contribution Boundaries}
\label{sec:related}

\subsection{Security-token standards}

ERC-1400 supplies controller operations and restriction metadata~\cite{erc1400}. ERC-3643 integrates identities, compliance, roles, recovery, forced transfer, and freeze controls~\cite{erc3643}. ERC-7943 deliberately specifies neutral RWA mechanics~\cite{erc7943}. ERC-1450 centralizes execution in a registered transfer agent and defines request lifecycle and regulatory operational patterns~\cite{erc1450}. Proposed ERC-8319 supplies the six legal-effect meanings that this paper mechanizes~\cite{erc8319}. The contribution is not a new collection of privileged functions; it is a reference semantics and evidence boundary for those meanings.

\subsection{Typed legal and business workflows}

Stipula gives legal contracts a formal semantics, observational equivalence, and type inference~\cite{stipula2023}. Daml provides typed choices, authorization, delegation, and atomic commit for real-world multiparty workflows~\cite{daml2023}. These systems foreclose any claim that legal meaning or authorization was first represented by types here. ERC-TRUST addresses a narrower Ethereum security-token execution vocabulary and its model-to-runtime evidence obligations.

\subsection{Smart-contract formal verification}

The closest literature includes Isabelle reasoning over EVM bytecode~\cite{amani2018}, executable EVM semantics and verification~\cite{kevm2018}, a deep embedding of Solidity in Isabelle/HOL~\cite{isabellesolidity2025}, and verified compilation to EVM~\cite{deepsea2024}. Compared with these systems, the current candidate is weaker in cross-layer proof strength. Its distinct value is the regulatory-action domain theory, explicit external-truth boundary, witness-and-mutation falsification, and evidence stratification for an independently written candidate.

\subsection{Mutation and provenance}

Solidity mutation testing and exact-match artifact verification are established techniques~\cite{sumo2021,sourcify}. The artifact applies them to make domain claims falsifiable and release identity inspectable.

\section{Conclusion}
\label{sec:conclusion}

This paper mechanizes a reference execution semantics for the six regulatory effects proposed in ERC-8319 and instantiates it for ERC-TRUST. Within the declared model, the artifact establishes typed action and reversal behavior, three-way outcomes, replay and epoch rules, complete frames, case-local terminality, canonical receipts, constructive nonvacuity, and a theorem-level boundary against inferring external legal truth. For one Solidity/EVM candidate, it records qualified current-profile obligations and separately scoped Foundry, Certora, Kontrol/KEVM, mutation, deterministic-build, and runtime-identity evidence.

The four research questions receive matching answers. RQ1 is answered by the typed semantics and its external-truth boundary (\cref{sec:semantics}); RQ2 by the mechanized theorem groups, their constructive witnesses, and their direct mutations (\cref{sec:mechanization,sec:countermodels}); RQ3 by the retrieve and transaction relations, the conditional profile-scoped theorems, and the named open central obligation (\cref{sec:implementation}); RQ4 by the evidence stratification and artifact-identity discipline that keep each result in its own class (\cref{sec:evidence,sec:artifact}).

The open obligations for public contributors are explicit: (i) a machine-checked construction of the runtime-link relation from source, compiler or bytecode semantics, external-call assumptions, and trace observations, discharging the conditional composition; (ii) a mechanized replacement for the trusted K/Isabelle word-correspondence record (\cref{sec:implementation}); (iii) a Verified Full ERC-3643 profile with atomic fresh deployment, a complete initial-state gate, a same-transaction transfer or Compliance hook, actual upstream post-state equality, and upgrade binding; (iv) a deployment manifest binding chain, address, runtime code, constructor arguments, roles, and dependencies for a specific deployment; (v) adversarial review of model adequacy through new counterexamples or new objection identifiers; and (vi) alternative conforming architectures, including proxy, migration, and batch profiles, evaluated against the same semantics with new conformance evidence. Appendix~\ref{app:correspondence} records the per-obligation status.

Three elements are reusable outside this standard: the typed outcome partition with complete persistent-state stutter for regulated assets, the evidence-stratification discipline that keeps heterogeneous tool results from combining into an unproved stronger theorem, and the stable-identifier objection register whose dispositions any reviewer can overturn with a stronger standard, theorem, trace, or counterexample.

The result is deliberately incomplete across layers. A conditional runtime-link locale is not a complete compiler-to-EVM proof. Tool PASS results do not add into such a theorem, and no deployment or legal truth is verified. This boundary is part of the contribution: it gives standard reviewers a stable target for stronger proofs, counterexamples, alternative architectures, and public objections without requiring them to accept a stronger claim than the artifact supports.

\appendix

\section{Theorem Inventory}
\label{app:theorems}

\Cref{tab:theorem-inventory} lists the load-bearing Isabelle declarations behind each manuscript result. It selects principal declarations rather than counting every helper lemma as a contribution. The exact sources are part of the release-manifest source root stated in \cref{sec:artifact}; theory files whose historical names embed internal milestone prefixes are described by role here and are recoverable by searching the repository for the stated theorem names.

\begingroup
\footnotesize
\begin{longtable}{@{}P{5.9cm}P{10.9cm}@{}}
\caption{Load-bearing declarations by manuscript result.}
\label{tab:theorem-inventory}\\
\toprule
Result and boundary & Source and principal declarations \\
\midrule
\endfirsthead
\toprule
Result and boundary & Source and principal declarations \\
\midrule
\endhead
\bottomrule
\endlastfoot
Typed action inventory and mapping. \emph{Boundary:} maps normative identities; does not prove legal truth. &
\path{RCP_Action_Mapping.thy}: \path{rcp_action_inventory}, \path{six_actions_are_not_seven_transition_labels}, \path{recover_and_liquidate_are_transfer_layer_actions}, \path{deescalation_labels_are_not_rcp_actions}, \path{forward_mapping_roundtrip} \\
\addlinespace
Canonical execution and outcome partition. \emph{Boundary:} abstract state and observation alphabet only. &
\path{Regulatory_Execution_Semantics.thy}: \path{every_entrypoint_converges_on_canonical_execution}, \path{untyped_entrypoint_is_fail_closed}, \path{canonical_rejection_is_full_state_stutter}, \path{canonical_operational_failure_is_full_state_stutter}, \path{successful_execution_complete_frame}, \path{successful_execution_exact_core_effects}, \path{successful_execution_writes_applied_receipt_last} \\
\addlinespace
Concrete nonvacuity and regulatory cases. \emph{Boundary:} witnesses do not by themselves establish model adequacy. &
\path{Regulatory_Execution_Simulation.thy}: action and outcome witnesses; \path{ce02_retrieve_relation}; \path{ce11_liquidate_assume_guarantee}; \path{ce11_liquidate_binds_settlement_without_asserting_external_truth}; \path{ce12_recover_assume_guarantee}; \path{ce12_recover_binds_entitlement_destination_and_consumption} \\
\addlinespace
Executable manifest evaluation. \emph{Boundary:} manifest rows are evaluated from reachable fixtures, not copied constants. &
\path{Executable_Regulatory_Kernel.thy}: \path{trust_manifest_domain_is_exactly_six_by_three}, \path{trust_manifest_kernel_is_actual_execution_evaluation}, \path{every_manifest_fixture_scenario_is_reachable}, \path{six_by_three_and_five_by_seven_are_not_conflated} \\
\addlinespace
External-truth nonclaim. \emph{Boundary:} an epistemic boundary, not the truth or falsity of any external world. &
\path{Claim_Boundary.thy}: \path{same_core_input_cannot_distinguish_external_legal_truth}, \path{external_worlds_really_disagree}, \path{ce02_guarantee_is_declared_custody_not_legal_title}, \path{ce11_guarantee_is_binding_not_sale_or_debt_truth}, \path{ce12_guarantee_is_binding_not_rightful_owner_truth}, \path{allowed_claim_inventory_is_exact} \\
\addlinespace
Token compatibility. \emph{Boundary:} the compatibility model does not prove arbitrary concrete deployments. &
\path{Token_Compatibility.thy}: \path{erc7943_frozen_amount_increase_converges_to_freeze}, \path{erc7943_frozen_amount_decrease_converges_to_unfreeze}, \path{erc7943_forced_transfer_requires_typed_binding}, \path{erc7943_forced_transfer_fails_closed_without_binding}, \path{erc3643_profile_is_an_explicit_subset}, \path{compatibility_does_not_expand_the_claim_boundary} \\
\addlinespace
Transaction-level abstract relation. \emph{Boundary:} an Isabelle relation over modeled concrete transactions. &
\path{TRUST_Transaction_Refinement.thy}: action success results, three reversal results, \path{rejection_is_persistent_stutter}, \path{operational_failure_is_persistent_stutter}, \path{malformed_transaction_is_abstract_stutter}, \path{dependency_revert_is_abstract_stutter}, \path{success_has_final_canonical_receipt_event} \\
\addlinespace
Current-state retrieve relation. \emph{Boundary:} the exact-runtime premise is part of the relation. &
\path{TRUST_Retrieve_Relation.thy}: \path{current_state_abstraction_well_defined}, \path{alpha_current_is_functional}, \path{alpha_current_requires_exact_runtime}, \path{alpha_current_rejects_runtime_substitution}, \path{nonce_projection_is_exact}, \path{case_terminality_is_scoped} \\
\addlinespace
Reusable package summaries. \emph{Boundary:} summaries consume package premises; they are not a new EVM semantics. &
\path{TRUST_Reusable_Summaries.thy}: \path{action_summary_success_commits_exact_abstract_state}, \path{action_summary_failure_restores_state_and_authorization}, \path{successful_summary_receipt_storage_return_and_final_event_agree} \\
\addlinespace
Conditional heterogeneous composition. \emph{Boundary:} the locale assumes \path{runtime_link}; the theorem named \path{end_to_end_refinement} is conditional. &
\path{TRUST_End_To_End_Composition.thy}: \path{compiler_correctness_remains_a_nonclaim}, \path{runtime_execution_implements_concrete_transaction}, action and reversal consequences, \path{runtime_rejection_stutters}, \path{runtime_operational_failure_stutters}, \path{end_to_end_refinement}, \path{heterogeneous_composition_is_conditional_until_certificate} \\
\addlinespace
Decoder guard composition. \emph{Boundary:} the cross-kernel correspondence record remains a trusted seam. &
Decode-slice and redundant-hypothesis theories: exact length, enumeration, and fixed-width unsigned guard results together with redundant-hypothesis theorems for the fixed 256-bit domain. \\
\addlinespace
Generated runtime bridge. \emph{Boundary:} generated identities bind declared compiled artifacts; they do not construct the runtime-link relation. &
\path{TRUST_Runtime_Bridge_Generated.thy}: \path{runtimes_are_within_the_eip170_limit}, \path{native_routes_are_exhaustive}, \path{profile_adapter_routes_are_exhaustive}, \path{profile_governor_routes_are_exhaustive}, \path{typed_command_selectors_are_classified_as_kernel_commands}, \path{generic_dispatcher_selector_is_unclassified} \\
\addlinespace
ERC-3643 Partial onboarding. \emph{Boundary:} declared-entry and touched-account behavior; no manifest completeness or ordinary-transfer hook theorem. &
\path{TRUST_Verified_Profile_Onboarding.thy}: \path{current_erc3643_reference_is_partial_not_verified_full}, \path{empty_manifest_changes_no_declared_state}, \path{forced_transfer_restriction_mismatch_is_an_operational_stutter}, \path{restriction_observation_reads_actual_upstream_flags}, \path{resynchronisation_never_lowers_an_owned_frozen_amount} \\
\addlinespace
Generated obligation ledger. \emph{Boundary:} names current evidence anchors and leaves the runtime link conditional. &
\path{TRUST_Obligation_Ledger_Generated.thy}: per-row named lemma bundles, \path{obligation_ledger_rows_are_distinct}, \path{obligation_ledger_has_no_current_mandatory_rows}, \path{obligation_ledger_binds_the_generated_bridge}, \path{obligation_ledger_closure_status_is_declared} \\
\addlinespace
Current-profile state distinctions. \emph{Boundary:} direct negative adequacy for two load-bearing distinctions. &
Current-profile state theories: \path{freeze_and_restriction_are_independent}, \path{state04_conflated_projection_is_distinguished}, \path{case_terminality_is_scoped}, \path{state05_global_terminal_mutant_is_distinguished} \\
\addlinespace
Current-profile row consequences. \emph{Boundary:} conditional on named package receipts and the product inhabitant. &
Row-corollary theories: per-row corollaries, action and reversal omission distinctions, \path{freeze_success_refines_if_current_profile_receipts_hold} \\
\end{longtable}
\endgroup

\section{Assumption and Nonclaim Ledger}
\label{app:assumptions}

\begin{table}[H]
\centering
\caption{Principal assumptions and nonclaims.}
\small
\begin{tabular}{@{}P{3.2cm}P{4.1cm}P{4.0cm}@{}}
\toprule
Item & Consumed by & Not established \\
\midrule
Runtime-link relation & Conditional composition theorems & Compiler-to-EVM correspondence \\
Bound dependency identity & Native action execution & Truth of the returned external fact \\
Pinned compiler and settings & Runtime identity & Compiler correctness \\
Pinned KEVM/Kore/solver stack & Selected symbolic claims & Tool implementation correctness \\
Declared ERC-3643 import entries and touched-account post-state & Partial adapter reference & Complete imported state, ordinary-transfer hook coverage, or arbitrary deployment conformance \\
Current-profile receipt set & Package and row corollaries & A single end-to-end proof object \\
No deployment input & Current claim boundary & Chain, address, constructor, roles, dependencies \\
\bottomrule
\end{tabular}
\end{table}

\section{Model-to-Code Correspondence Map}
\label{app:correspondence}

\Cref{tab:correspondence} records, for every semantic obligation, its abstract anchor, its concrete anchor in the candidate, the evidence that currently exists, and the obligation that remains open. The open-obligation column is the working map for public contributors.

\begingroup
\footnotesize
\begin{longtable}{@{}P{2.1cm}P{3.3cm}P{3.3cm}P{3.6cm}P{3.8cm}@{}}
\caption{Semantic obligations, present evidence, and open obligations.}
\label{tab:correspondence}\\
\toprule
Obligation & Abstract anchor & Concrete anchor & Present evidence & Open obligation \\
\midrule
\endfirsthead
\toprule
Obligation & Abstract anchor & Concrete anchor & Present evidence & Open obligation \\
\midrule
\endhead
\bottomrule
\endlastfoot
Six typed actions & Action datatype, mapping, success semantics & Action dispatcher and effect storage in the token source & Unit/fuzz tests, financial-core CVL rules, current-profile row corollaries & One proof object deriving every runtime path from the abstract step relation \\
\addlinespace
Three reversals & Reversal type, original-action reference, admissibility & Reversal dispatcher and prior-state restoration & Unit tests, reversal CVL package, row corollaries & Full correspondence for all failure, storage, and event paths \\
\addlinespace
Outcome partition & Abstract three-way partition and full-state stutter & Custom errors, revert behavior, dependency classifier & Foundry, CVL, selected KEVM/Kontrol seams & Ambient observations such as gas and mempool behavior remain outside the model \\
\addlinespace
Authorization, epoch, nonce, replay & Command lifecycle and stale invalidation & Action identifiers, authority and delegation epochs, nonce consumption & Authorization/replay package, mutation, tests & Authority legitimacy and key operations remain external \\
\addlinespace
External dependency binding & Bound address, code, config, schema, epoch, input & Read-only fail-closed policy and evidence adapters & Classifier CVL, malformed-return and revert seams & Truth of policy, identity, settlement, proceeds, or entitlement is not established \\
\addlinespace
ERC-7943 exact-use path & Strict-increase FREEZE, explicit UNFREEZE, and fail-closed untyped path & Same-transaction route ticket and nonincreasing-target guard & Route invariant, raw-selector and failure-stutter symbolic evidence, direction and deletion mutants & Arbitrary third-party implementations require their own inventory \\
\addlinespace
ERC-3643 profile & Explicit optional subset & Partial adapter plus sealed governor & Declared-entry import checks, touched-account post-state tests, four Partial CVL rules, runtime identity & Complete initial state and same-transaction ordinary-transfer enforcement require a new hook-enabled deployment profile \\
\addlinespace
Success effects and receipt & Exact write set, final receipt observation & Regulatory state owner, effect record, receipt storage and event & Current-profile packages, tests, selected final-log proof & One compiled-runtime simulation covering every effect and emitted trace \\
\addlinespace
Runtime identity & Exact-runtime retrieve relation & Pinned compiler, ABI, storage, creation/runtime bytecode, method IDs, immutable references & Deterministic build, runtime-binding verifier, negative classification campaign & Compiler correctness and deployed-address identity \\
\addlinespace
Deployment & Intentionally outside the current model & No deployed address in the candidate evidence & None claimed & A future deployment manifest must bind chain, address, runtime, constructor, roles, and dependencies \\
\end{longtable}
\endgroup

\section{Tool-Evidence Matrix}
\label{app:tools}

\begin{table}[H]
\centering
\caption{Evidence classes and their ceilings.}
\small
\begin{tabular}{@{}P{2.2cm}P{4.5cm}P{4.5cm}@{}}
\toprule
Layer & Establishes & Does not establish \\
\midrule
Isabelle/HOL & Theorems in the declared mathematical domain & Solidity, compiler, EVM, or deployment correctness \\
Certora & Configured CVL rules under harness and summary assumptions & Full model correspondence or all external environments \\
Foundry & Concrete unit, fuzz, invariant, and mutation evidence & Exhaustive state-space soundness \\
Kontrol/KEVM & Selected compiled-EVM path and seam claims & Every runtime path or the missing central relation \\
Manifests/build & Provenance and declared artifact identity & Compiler correctness \\
Mutation & Sensitivity to declared faults & Complete fault model or proof adequacy \\
Deployment manifest & One deployment's identity if later supplied & Legal truth or universal operational safety \\
\bottomrule
\end{tabular}
\end{table}

\section{Falsification Inventory}
\label{app:falsification}

The declared families include action/receipt identity removal, stale-epoch acceptance, outcome conflation, frame removal, global terminality, reversal prior-state omission, Native and adapter nonincreasing-FREEZE acceptance, exact-use ticket deletion, custody-accounting omission, receipt-binding omission, semantic runtime drift, packaging-only drift, and expected-hash overwrite. Each public mutation must identify its consumer; an unconsumed mutation does not earn evidence credit. \Cref{tab:falsification} summarizes the counted campaigns at the pinned candidate; counts are recounted at the merged public commit before release.

\begin{table}[H]
\centering
\caption{Counted falsification campaigns at the pinned candidate.}
\label{tab:falsification}
\small
\begin{tabular}{@{}P{4.2cm}P{3.2cm}P{3.4cm}P{4.6cm}@{}}
\toprule
Campaign & Layer & Result & Declared consumer \\
\midrule
Abstract model mutations & Isabelle model & 15 of 15 detected & Consumer theorems and closure checks \\
Implementation mutations & Solidity candidate & 121 of 121 killed & Foundry mutation harness and obligation-ledger consumers \\
Runtime-binding verifier self-mutations & Artifact binding & 18 of 18 killed & Runtime-binding verifier \\
Independent specification reproduction & Generated kernel surface & 23 vectors, 401 assertions PASS & Independent hand-written reproducer \\
\bottomrule
\end{tabular}
\end{table}

\section{Claim-to-Source Map}
\label{app:claims}

\Cref{tab:claim-map} assigns every headline statement to its evidence class, its required qualifier, and the expansion it must never receive. The abstract, introduction, contribution summary, and conclusion all repeat the absence of a complete end-to-end theorem, compiler correctness, audit, production readiness, deployment verification, and external legal-truth proof.

\begingroup
\footnotesize
\begin{longtable}{@{}P{4.6cm}P{3.3cm}P{4.2cm}P{4.4cm}@{}}
\caption{Headline claims, evidence classes, qualifiers, and forbidden expansions.}
\label{tab:claim-map}\\
\toprule
Claim & Evidence class & Required qualifier & Forbidden expansion \\
\midrule
\endfirsthead
\toprule
Claim & Evidence class & Required qualifier & Forbidden expansion \\
\midrule
\endhead
\bottomrule
\endlastfoot
Six regulatory actions and three reversals have distinct typed identities and effects in the abstract model & Machine-checked theorem & Within the declared Isabelle/HOL semantic domain & Legal validity or correct real-world classification \\
\addlinespace
Applied, Rejected, and OperationalFailure are distinct outcomes; the latter two preserve the declared persistent state & Machine-checked theorem & Persistent abstract state and declared observation alphabet & Gas, timing, mempool, or every environmental observation \\
\addlinespace
Replay protection and stale-authorization invalidation are modeled by bound identifiers, epochs, nonce consumption, and cancellation & Theorem plus bounded implementation evidence & Declared request domain and current candidate & Authority legitimacy, key custody, or universal replay freedom for every integration \\
\addlinespace
Action and reversal success paths are nonvacuous & Constructive model witness and concrete tests & Witnessed declared paths & Complete model adequacy or production reachability \\
\addlinespace
External legal and factual truth is deliberately unprovable from the kernel's bound inputs alone & Machine-checked indistinguishability boundary & Same kernel input can correspond to different external worlds & A claim that external attestations are false or useless \\
\addlinespace
The native reference candidate implements the six typed actions and three reversals & Source inspection and bounded tool evidence & Exact candidate and declared current profile & Whole-program or arbitrary-deployment correctness \\
\addlinespace
The current successor profile has twelve named evidence lanes PASS with no pending lane; the preserved package projection remains 7/7 reusable packages, 49/49 Core, 24/24 mandatory Supporting, and 0/6 optional Verified Full & Hash-bound release evidence & Distinct projections with no partial credit; ledger closure remains conditional & A complete end-to-end refinement theorem or third-party certification \\
\addlinespace
The conditional Isabelle composition derives action-specific results from a runtime-link premise & Machine-checked conditional theorem & \path{runtime_link} is a locale assumption & Describing \path{end_to_end_refinement} as a discharged compiler-to-EVM proof \\
\addlinespace
Selected ABI, call, revert, rollback, log, receipt, decoder, and runtime-identity seams have symbolic or bounded evidence & Tool-specific bounded or symbolic evidence & Only the named seam, source, schedule, and assumptions & All EVM paths, all external contracts, or compiler correctness \\
\addlinespace
Deterministic builds and runtime binding identify the tested artifact & Provenance and negative identity evidence & Identical declared inputs and pinned toolchain & Compiler correctness or deployed-address identity \\
\addlinespace
Mutation campaigns show that declared fault operators affect their intended consumers & Negative adequacy evidence & Only the declared mutation family & Completeness of the fault model or proof soundness by mutation score \\
\addlinespace
The ERC-3643 reference is Partial and never Full & Bounded conformance evidence & Declared import entries, touched-account checks, sealed governor, and exact current runtime only & Manifest completeness, same-transaction ordinary-transfer enforcement, or arbitrary-deployment conformance \\
\end{longtable}
\endgroup

\section{Verification Commands}
\label{app:commands}

All commands run from the root of the artifact repository~\cite{trustartifact} at the commit stated in \cref{sec:artifact}. The scripts are PowerShell; on Linux or macOS, install PowerShell~7 and replace \isaf{powershell -File} with \isaf{pwsh -File}. Every layer runs locally except the Certora layer, which requires a Certora account. Each command establishes only its own layer's facts, so the commands remain separate by design.

\begingroup
\footnotesize
\begin{verbatim}
# Full current-profile replay and its verifier
powershell -File scripts/replay-current-profile-release.ps1
node scripts/verify-current-profile-release-v3.mjs

# Isabelle model closure and model-level negative mutations
powershell -File formal/isabelle/ERC_TRUST/evidence/model-verification/run-trust-closure.ps1
powershell -File formal/isabelle/ERC_TRUST/evidence/model-verification/run-negative-mutations.ps1

# Solidity build (with runtime sizes) and Foundry tests
forge build --sizes
forge test

# Current ERC-3643 Partial Certora rules (requires a Certora account)
certoraRun implementation/certora/ERC3643Partial.conf

# Kontrol/KEVM claims (four names in evidence/kontrol-results-v3.json)
kontrol prove --foundry-project-root . --schedule CANCUN --match-test <proof-name>

# Artifact identity: deterministic build, release identity, runtime binding
powershell -File scripts/check-deterministic-build.ps1
node scripts/verify-release.mjs
node scripts/generate-runtime-binding-v3.mjs --check
node scripts/verify-runtime-binding-v3.mjs --replay
\end{verbatim}
\endgroup

The proof-name placeholder enumerates the four entries in \path{evidence/kontrol-results-v3.json}. The repository scripts and configuration files are the executable source of truth.

\section{Full Objection-Disposition Table}
\label{app:objections}

\Cref{tab:objections} lists all 22 objections under stable identifiers. Every objection is stated in its strongest plausible form; initial dispositions are manuscript decisions and may change after public review. Public review may add new identifiers without renumbering existing ones.

\begingroup
\footnotesize
\begin{longtable}{@{}P{0.95cm}P{4.45cm}P{5.1cm}P{3.4cm}P{2.3cm}@{}}
\caption{Objections, dispositions, and residual limits.}
\label{tab:objections}\\
\toprule
ID & Strongest objection & Evidence and response & Residual limit & Disposition \\
\midrule
\endfirsthead
\toprule
ID & Strongest objection & Evidence and response & Residual limit & Disposition \\
\midrule
\endhead
\bottomrule
\endlastfoot
01 & ERC-3643 separates recovery and freezing, ERC-1450 supplies a request lifecycle, ERC-7943 defines neutral mechanics; TRUST merely recombines existing controls & The paper does not deny those typed controls. Its narrower increment is one common semantics for all six ERC-8319 legal effects, especially effects sharing one balance-movement mechanism while requiring different custody, settlement, entitlement, terminality, and receipt obligations & Adoption value is a standards question, not a theorem & bounded \\
\addlinespace
02 & A contract cannot decide legal truth, so typed regulatory execution is conceptually invalid & The model proves the separation: identical kernel input can coexist with disagreeing external worlds. Applied means the bound on-chain transition occurred, not that a court order, title, sale, debt, or entitlement is true & Bad or fraudulent external input can still drive verified code & accepted limitation \\
\addlinespace
03 & Six actions freeze jurisdiction-specific policy into one taxonomy & The action vocabulary fixes observable effect classes; jurisdiction policy remains in versioned external bindings and authority configuration & Some jurisdictions may require additional effect classes or profiles & bounded \\
\addlinespace
04 & Freeze/unfreeze and pause/unpause already separate actions from reversals & The paper claims only the domain-specific combination of referenced prior state, stale, duplicate, and out-of-order rejection, case-local terminality, and typed receipt identity & The current domain declares only three reversals & fixed in declared domain \\
\addlinespace
05 & Typed expected-error and exceptional-failure categories predate TRUST, so the three-way partition is not new & The paper claims the regulatory audit and retry meanings plus complete declared-state and log stutter, not the invention of typed outcomes & Revert payload and monitoring infrastructure must preserve the distinction operationally & bounded \\
\addlinespace
06 & ERC-2612 and ERC-3009 already provide domain binding, nonce, deadline, cancellation, and single consumption & TRUST combines those established controls with action-specific case, epoch, evidence, and receipt semantics; it does not claim to invent replay protection & Key compromise and authority legitimacy remain outside the proof & bounded \\
\addlinespace
07 & The exact-use ticket is only another one-time authorization pattern & Its narrow increment binds a legacy selector, caller, calldata, policy, epochs, and command to same-transaction immediate consumption; raw ERC-7943 selectors remain closed in the candidate & Third-party token variants need separate topology evidence & fixed for candidate \\
\addlinespace
08 & ERC-3643 owner, Agent, batch, ordinary-transfer, or compliance paths can bypass an adapter & The current descriptor is explicitly Partial and never Full. It checks declared imports, touched-account post-state, actual restriction observations, and sealed governance without claiming a complete initial-state inventory or ordinary-transfer hook & Verified Full requires a hook-enabled fresh deployment and a complete initial-state gate & accepted limitation \\
\addlinespace
09 & Isabelle and Solidity/EVM remain separated by an assumed correspondence & Correct. The paper exposes the conditional locale, retrieve relation, package receipts, and selected runtime seams without calling them a complete central theorem & Full compiler-to-runtime simulation remains open & accepted limitation \\
\addlinespace
10 & Multiple tools do not compose into soundness & Each tool receives its own projection, and upward inference from combined PASS counts is forbidden & Tool implementations and configurations remain in the trusted computing base & fixed in claim discipline \\
\addlinespace
11 & Fuzz and invariant counts create a false impression of generality & They are reported as bounded counterexample evidence with exact runs and calls & Unexplored sequences and inputs remain possible & bounded \\
\addlinespace
12 & Mutation kills do not establish proof adequacy & Mutations demonstrate sensitivity only for declared fault operators and consumer claims & The fault model is not complete & accepted limitation \\
\addlinespace
13 & Nonvacuity witnesses can validate a wrong model against itself & Abstract witnesses are paired with concrete positive and negative paths and selected EVM seams & Model adequacy still depends on specification review and external counterexamples & accepted limitation \\
\addlinespace
14 & Revert stutter cannot preserve gas, timing, mempool, or every observable side effect & The observation alphabet is limited to persistent declared state, declared downstream state, logs, and return/revert classification & Ambient execution observations are out of scope & fixed by scope \\
\addlinespace
15 & Hash binding is useless if the pinned compiler miscompiles & Provenance and bytecode evidence are separated, and compiler correctness is an explicit nonclaim & The pinned compiler is in the trusted computing base & accepted limitation \\
\addlinespace
16 & A verified candidate may differ from the deployed contract & This version makes no deployment claim. A future deployment manifest must bind chain, address, runtime, constructor, roles, and dependencies & No deployment evidence exists in this version & accepted limitation \\
\addlinespace
17 & A malicious or compromised authority can use verified code to execute wrongful actions correctly & Scope, epoch, cancellation, and receipt properties constrain execution & Legitimacy, governance quality, key custody, and lawful instruction are outside the theorem & accepted limitation \\
\addlinespace
18 & SEIZE custody and declared prior-holder fields do not restore legal ownership & The theorem deliberately guarantees custody accounting and declared relations, not beneficial title & External title adjudication remains outside the system & accepted limitation \\
\addlinespace
19 & A globally terminal implementation could block unrelated assets after one case closes & Current-profile state theorems make terminality case-scoped and distinguish a global-terminal mutant & Product changes must rerun this negative & fixed for candidate \\
\addlinespace
20 & Runtime growth can cross EIP-170 and invalidate a release & Two isolated builds report Native 20,043 bytes (margin 4,533), adapter 19,480 (margin 5,096), and governor 2,787 (margin 21,789), all bound to the current artifacts & Compiler or feature drift can consume the margins and must rerun the gate & bounded \\
\addlinespace
21 & ERC-8319 is not canonical, so TRUST cannot claim a canonical dependency & The manuscript says proposed ERC-8319, records its pending status, and states that TRUST has no assigned number and no opened pull request & Number and dependency status are restated in any future revision & fixed in provenance \\
\addlinespace
22 & The manuscript is a standards advertisement rather than an academic contribution & The paper separates the domain execution semantics, its theorem and falsification structure, the external-truth impossibility boundary, the evidence taxonomy, and the open correspondence obligations from TRUST-specific advocacy & Novelty judgment ultimately rests with reviewers and future prior art & bounded \\
\end{longtable}
\endgroup

\section{Terminology}
\label{app:terminology}

\begin{description}[leftmargin=1.2em,itemsep=1pt,parsep=1pt]
  \item[Case.] An identifier for one regulatory proceeding, such as one seizure or liquidation matter. Terminality is declared per case, never for the whole token.
  \item[Core observation.] The artifact's projection of the kernel state and entry that canonical execution may depend on; external-world fields are deliberately excluded from it.
  \item[CVL.] Certora Verification Language, the rule language checked by the Certora Prover.
  \item[Distinguishing negative.] A mutated artifact or projection that the current profile's checks must reject, demonstrating that the check can fail.
  \item[Evidence package.] A named group of expensive verification results, such as authorization and replay, shared by several obligation rows through hash-bound certificates.
  \item[Exact-use ticket.] A one-time authorization record bound to the exact caller, selector, calldata, policy binding, epochs, and command identifier, created and consumed within a single transaction.
  \item[Executable manifest.] The evaluated action-by-outcome scenario table computed from the executable kernel semantics rather than written down as constants.
  \item[Frozen target.] The absolute frozen-token amount recorded for a holding; FREEZE must strictly increase it, while a decrease requires an explicit UNFREEZE reversal. It is not a relative delta or the holder's address.
  \item[Hash-bound.] Valid only for artifacts whose cryptographic hashes equal those recorded in the binding record.
  \item[Kernel.] The immutable execution core that evaluates typed commands and owns regulatory state and receipts.
  \item[Locale.] An Isabelle mechanism for a named context of fixed parameters and assumptions; theorems inside a locale are conditional on its assumptions.
  \item[Native path.] The kernel's own execution path, as opposed to the optional ERC-3643 adapter profile.
  \item[Nonvacuity.] The property that a theorem's premises are satisfiable, so the theorem is not true merely because no state meets them.
  \item[Obligation row.] One entry in the published current-profile index; each row carries its own evidence and qualification certificate.
  \item[Packaging drift.] A change confined to artifact wrapping or metadata that leaves the semantic payload byte-identical; semantic drift, by contrast, changes the payload.
  \item[Publication profile.] The declared, versioned scope of obligations and configurations under which evidence is claimed for the exact candidate.
  \item[Qualification certificate.] A machine-checkable record stating that a row's or package's evidence was produced against artifacts with exactly the recorded hashes. Qualification is a published, repository-defined pass criterion, not a third-party certification, audit, or regulatory approval.
  \item[Retrieve relation.] The abstraction mapping from a concrete storage state to an abstract model state, in the sense used by refinement methods.
  \item[Runtime link.] The assumed relation stating that an accepted runtime execution satisfies the abstract transaction relation. Constructing it is the open central obligation.
  \item[ERC-3643 Partial reference.] The current interoperability adapter checks declared import entries and every account it touches, binds actual restriction observations, and uses sealed governance, while explicitly leaving manifest completeness and ordinary-transfer hook coverage unproved. Its descriptor is \texttt{PARTIAL/full=false}.
  \item[Seam.] A narrow, high-risk boundary in the compiled EVM code, such as ABI decoding or external-call classification, verified in isolation rather than by whole-program proof.
  \item[Stutter.] A step that leaves every declared persistent field unchanged; the term is borrowed from refinement theory.
  \item[Witness.] A concrete state and input on which a scenario actually executes, showing that a premise is satisfiable.
\end{description}

\paragraph{Disclosure.} The author is a co-author of proposed ERC-8319 and leads the development of the ERC-TRUST reference candidate at Oraclizer Labs. The claim-to-source map, the stable objection register, and the explicit nonclaims exist partly so that this interest can be checked against the artifact rather than taken on trust.

\paragraph{License.} This manuscript is licensed under the Creative Commons Attribution 4.0 International License (CC BY 4.0). Software, tests, and machine-checkable artifacts in the repository are licensed under BSD 3-Clause; the pre-ERC specification text is dedicated under CC0 1.0, matching the EIP-1 copyright requirement for an eventual ERC submission.

\begingroup
\small
\bibliographystyle{plain}
\bibliography{references}
\endgroup

\end{document}